\documentclass[aps,prx,reprint,amsmath,amssymb,superscript address,notitlepage]{revtex4-2}
\usepackage{graphicx}% Include figure files

\usepackage{dcolumn}% Align table columns on decimal point
\usepackage{bm}% bold math
\usepackage{xcolor}
\usepackage{soul}
\usepackage{textcomp}
\usepackage{gensymb}
\usepackage{url}
\usepackage[plainpages=false,pdfpagelabels=true,linkcolor=blue,colorlinks=true]{hyperref}
\usepackage[version=4]{mhchem}
\begin{document}

\title{Quantum Oscillations and Superconductivity in YPtBi Under Pressure}

\author{Jared Z. Dans}
\author{Prathum Saraf}
\author{Lillian Jirousek}
\author{Carsyn L. Mueller}
\affiliation{Maryland Quantum Materials Center and Department of Physics, University of Maryland, College Park, MD, 20742, USA}

\author{Chandra Shekhar}
\affiliation{Max Planck Institute for Chemical Physics of Solids, 01187 Dresden, Germany}

\author{Claudia Felser}
\affiliation{Max Planck Institute for Chemical Physics of Solids, 01187 Dresden, Germany}
\affiliation{Canadian Institute for Advanced Research, Toronto, Ontario, Canada}

\author{Johnpierre Paglione}
\email{paglione@umd.edu}
\affiliation{Maryland Quantum Materials Center and Department of Physics, University of Maryland, College Park, MD, 20742, USA}
\affiliation{Canadian Institute for Advanced Research, Toronto, Ontario, Canada}

% \date{28 February, 2026}
\date{\today}

\begin{abstract} % USE FOR REVTEX
    The topological semimetal YPtBi has attracted considerable attention, owing to its novel superconducting and normal state properties. A strong band inversion from spin-orbit coupling allows the existence of $j=3/2$ quasiparticles near the Fermi level, which form Cooper pairs with angular momentum potentially higher than single or triplet states. In this report, we present high-pressure magnetotransport and Shubnikov-de Haas effect measurements on high-quality YPtBi up to $P = 2.08$ GPa. As a function of pressure, we observe a trend toward more insulating resistivity at low temperatures concomitant with a suppression of quantum oscillation amplitude. Together with a decrease of the upper critical field and significant increase in the Dingle temperature, the pressure-induced changes point to a weakening of the band inversion and potential tuning of the topological nature of YPtBi, suggesting pressure as a useful tool for understanding the nature of topology in other related half-Heusler compounds.
\end{abstract}

\maketitle

\section{Introduction}
The non-centrosymmetric Half-Heusler family is a home to a wide variety of materials that host exotic phases \cite{GrafHHReview}.
In particular, the RTBi series (R = rare earth, T = Pd,Pt) has emerged as an exciting platform to study the interplay between topological phases and other collective phases, such as superconductivity and magnetism \cite{NakajimaRPdBi, ButchYPtBi}.
Here we focus on the superconducting state of the topological semimetal YPtBi.

YPtBi undergoes a superconducting transition at just below 1 K \cite{ButchYPtBi} and has a very low carrier density of ${\sim}2\times 10^{18}$ cm$^{-3}$ \cite{ButchYPtBi, KimYPtBiBeyondTriplet, KimYPtBiQuantumOsc}.
This carrier density is roughly three orders of magnitude too low to support conventional superconductivity in a 1 K superconductor \cite{MeinertYPtBi}.
Furthermore, YPtBi exhibits an upper critical field of $H_{c2}(0) = 1.5$ T with $T$-linear behavior over the superconducting range, in contrast to the parabolic behavior expected for a conventional superconductor \cite{ButchYPtBi}.
Likewise, penetration depth measurements indicate nodes in the superconducting gap \cite{KimYPtBiBeyondTriplet, KimYPtBiCampbell}.
There are also hints of possible surface conduction and superconductivity from the angle dependence of the magnetoresistance and $H_{c2}$ \cite{KimYPtBiSurface}.
All together, this situates YPtBi as an exciting platform for unconventional superconductivity.

This still leaves question of the pairing mechanism, in which the topologically non-trivial band structure likely plays a role.
Band structure calculations suggest the strong spin-orbit coupling from Bi causes band inversion between the $s$-like $\Gamma_6$ band and $p$-like $\Gamma_8$ band with quadratic band-touching at the Fermi level \cite{FengHHBandStruc, LinHHBandStruc, ChadovHHBandStruc}, which makes YPtBi a zero-gap semiconductor or semimetal.
This can allow $j=3/2$ quasi-particles to exist near the Fermi level from the combination of $s=1/2$ spin angular momentum and the $l = 1$ orbital angular momentum from Bi $p$-orbitals.
This $j=3/2$ character was later confirmed by quantum oscillation (QO) measurements  that showed large amplitude variations depending on the direction of applied field, which has a natural explanation with a $j=3/2$ Fermi surface \cite{KimYPtBiQuantumOsc}.
The unusual Fermi surface opens up the possibility of $j=3/2$ quasiparticles forming Cooper pairs expanding the possible pairing states from the typical singlet and triplet ($J=0,1$) to quintet ($J=2$) and septet ($J=3$) pairing states \cite{KimYPtBiBeyondTriplet, BrydonPairing, VenderbosPairing, SavaryPairing, IshiharaPairing}.
These exotic pairing states could realize topological superconductivity when combined with the topologically non-trivial band structure \cite{BrydonPairing, VenderbosPairing}.

A key parameter for probing and tuning the electronic structure of unconventional superconductors is the application of hydrostatic pressure.
In other low carrier density superconductors, such as \ce{SrTiO3} and \ce{Sr_xBi2Se3}, a strong suppression of superconductivity has been observed under pressure \cite{PfeifferSrTiO3Pressure, NikitinSrBi2Se3Pressure}.
Meanwhile, the effect of pressure on YPtBi has remained somewhat underexplored.
One study was done early on, which reported an enhancement of $T_c$ and $H_{c2}$ in metallic YPtBi under pressure \cite{BayYPtBiPressure}.
However, this report did not include any quantum oscillation measurements.
Furthermore, high-quality semimetallic samples with much lower carrier densities have since been grown, and these newer generation samples have not been studied under pressure.

In this work, we investigate the effect of pressures up to 2.08 GPa on the Fermi surface and superconducting state in high-purity YPtBi.
We present pressure-dependent Shubnikov-de Haas (SdH) quantum oscillation measurements which show minimal changes in frequency and effective mass but a large increase in the scattering rate.
Low-temperature resistivity measurements show virtually no change $T_c$ with pressure; however, we do observe a broadening of the transition and a suppression of $H_{c2}$. % Mention band inversion/surface states?

\section{Experimental Methods}
High-quality single crystals of YPtBi were grown from excess Bi flux. A 1:1:20 ratio of 99.99\% Y, 99.999\% Pt, and 99.9999\% Bi were placed in an alumina crucible and sealed under argon in a quartz ampoule.  
The ampoule was held at 1100$^\circ$C for 12 hours and then slowly cooled to 520$^\circ$C before centrifuging to remove excess flux.
A sample was cut from a larger crystal and polished into a long, thin bar.
Gold wires were then attached using silver epoxy to enable four-point resistance measurements.

High-pressure transport experiments were performed in a piston-cylinder pressure cell using Daphne 7575 oil as the pressure medium.
A coil of manganin and a strip of Pb were used as room- and low-temperature manometers, respectively.
Magneto-transport measurements were performed in a Quantum Design PPMS equipped with a 14 T magnet from 2 - 300 K.
Sub-Kelvin resistivity measurements were performed using an AC resistance bridge in an Oxford Heliox $^3$He refrigerator.

\begin{figure*}[t]
    \centering
    \includegraphics[width=1\linewidth]{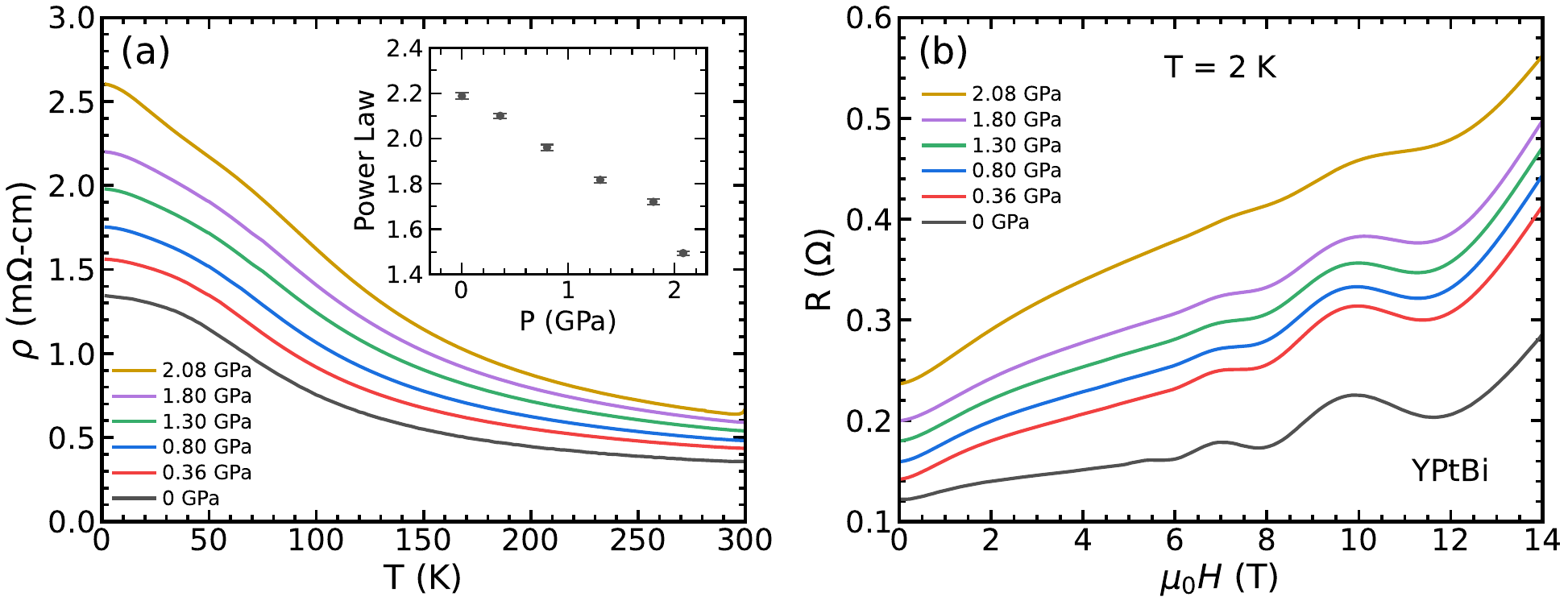}
    \caption{Evolution of transport behavior in YPtBi under pressure. (a) Resistivity as a function of temperature at selected pressures. Empirical power law fits to Eq. \ref{eq:PowerLaw} were performed below 100 K (inset) to characterize the evolution of the resistivity. (b) Low temperature magnetoresistance under pressure. Large Shubnikov-de Haas oscillations can be seen at all pressures.}
    \label{fig:Transport}
\end{figure*}

\begin{figure*}[t]
    \centering
    \includegraphics[width=1\linewidth]{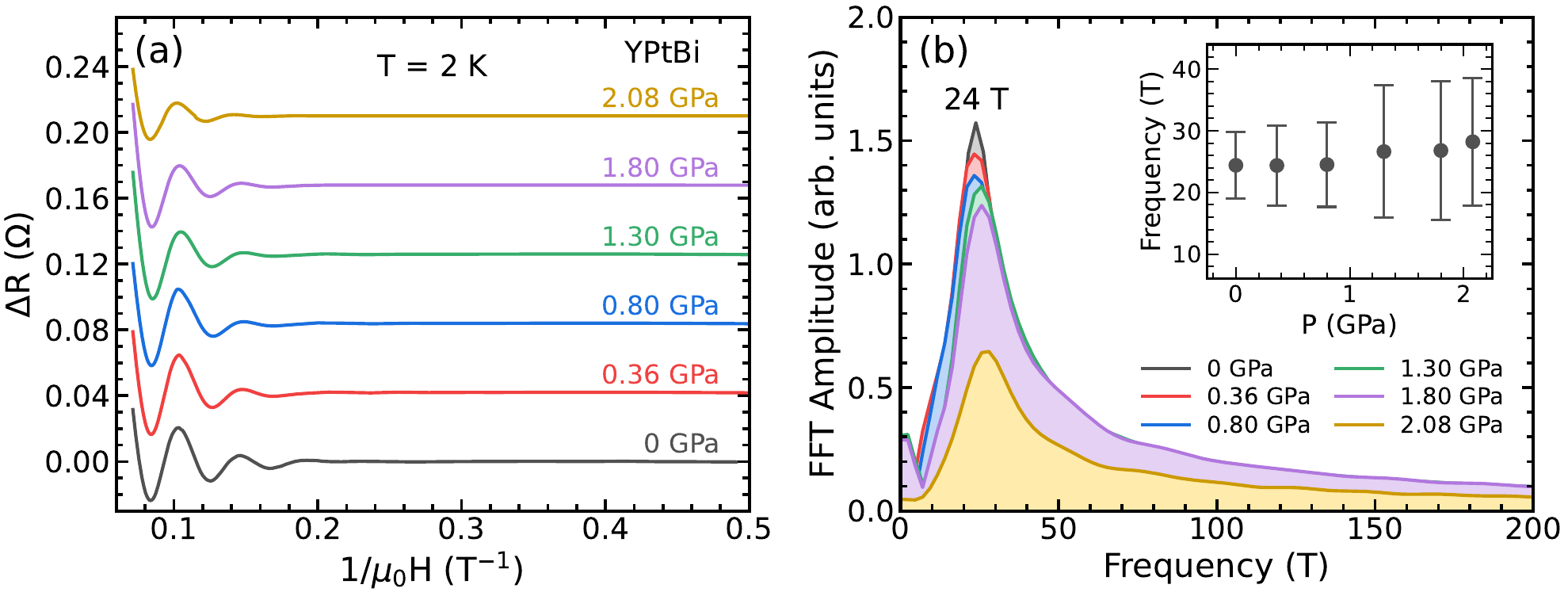}
    \caption{Pressure-dependent Shubnikov-de Haas quantum oscillations. (a) Oscillations $\Delta R$ extracted from the magnetoresistance. Under pressure, the oscillations show similar amplitudes at high field, but show stronger damping. (b) The pressure dependence of the oscillation frequency (inset) was determined via FFT (main). The oscillation frequency is largely unchanged by pressure.}
    \label{fig:QO_Pressure}
\end{figure*}

\section{Results and Discussion}
To begin, we first characterize the resistivity of YPtBi as a function of pressure.
Figure \ref{fig:Transport}(a) shows the resistivity versus temperature from 0 to 2.08 GPa.
At ambient pressure, the resistivity increases with decreasing temperature before saturating around 50 K.
As the pressure is increased, the resistivity gets larger at all temperatures and takes on a more insulating character.
To characterize the pressure-evolution of the low-temperature resistivity, we perform an empirical power law fit borrowed from a report on the semiconductor \ce{PtAs2} \cite{WangPtAs2Pressure}.
We fit the data to
\begin{equation}
    \rho(T) = \frac{1}{\sigma_0+aT^n}
    \label{eq:PowerLaw}
\end{equation}
where $\sigma_0$ is the residual low temperature conductivity, $a$ is a scaling coefficient, and $n$ is the power law exponent. 
As shown in the inset, the exponents extracted from the fit decrease monotonically from 2.2 to 1.5 with increasing pressure.
Like in Ref. \cite{WangPtAs2Pressure}, we use these fits as a simple way to characterize the pressure evolution, which is possibly complicated by both bulk and surface conductivity components that may evolve differently \cite{KimYPtBiSurface}.
Without further experiments to quantify these contributions, we keep the interpretation of the fits to a minimum.
Figure \ref{fig:Transport}(b) shows the magnetoresistance at the selected pressures. 
YPtBi exhibits a larger MR under pressure, and large Shubnikov-de Haas quantum oscillations are observed at all pressures.

Figure \ref{fig:QO_Pressure} shows SdH oscillations as a function of pressure.
The oscillations presented in Fig. \ref{fig:QO_Pressure}(a) were extracted from the magnetoresistance.
For pressures up to 1.80 GPa, the oscillation amplitudes are roughly the same at high field but are damped more heavily at low field as the pressure increases.
For $P = 2.08$ GPa, the amplitude is smaller across the entire field range.
This is reflected in the FFTs (Fig. \ref{fig:QO_Pressure}(b)), where small decreases in amplitude are seen up to 1.80 GPa, with a large decrease in amplitude at 2.08 GPa. 
The oscillation frequencies were extracted from the FFTs and plotted as a function of pressure in the inset.
The ${\sim}24$ T oscillation frequency is largely unchanged by pressure.
We calculate the carrier densities as $n_{0\ \mathrm{GPa}} = (6.8\pm2.3)\times10^{17}$ cm$^{-3}$ and $n_{2.08\ \mathrm{GPa}} = (8.5\pm4.7)\times10^{17}$ cm$^{-3}$ for 0 and 2.08 GPa, respectively. 
This is nearly two orders of magnitude smaller than reported by Bay et al. \cite{BayYPtBiPressure}.
Even neglecting the relatively large uncertainties, the carrier density does not vary much with pressure, consistent with minimal changes in the Fermi surface.

We next address the effect of pressure on the oscillation amplitudes.
The QO amplitude depends strongly on the effective mass, $m^*$, and the impurity scattering time, $\tau$.
In a semiclassical picture, QOs are only observable when $\omega_c\tau<1$, where $\omega_c = e\mu_0H/m$ is the cyclotron frequency.
In the Lifshitz-Kosevitch theory \cite{LKTheory}, the oscillatory part of the magnetoresistance is proportional to $\Delta R \propto(\mu_0H)^{1/2}A_T(T,H)A_D(H)$, where 
\begin{equation}
    A_T = \frac{\alpha T/\mu_0H}{\sinh(\alpha T/\mu_0H)}
    \label{eq:AT}
\end{equation}
\begin{equation}
    A_D = \exp\left(\frac{\alpha T_D}{\mu_0H}\right)
    \label{eq:AD}
\end{equation}
with $\alpha = 2\pi^2k_Bm^*/e\hbar$ and the Dingle temperature $T_D = \hbar/2\pi k_B\tau$. 
These parameters can then be extracted from the $T$- and $H$-dependence of the oscillations.

\begin{figure*}[t]
    \centering
    \includegraphics[width=1\linewidth]{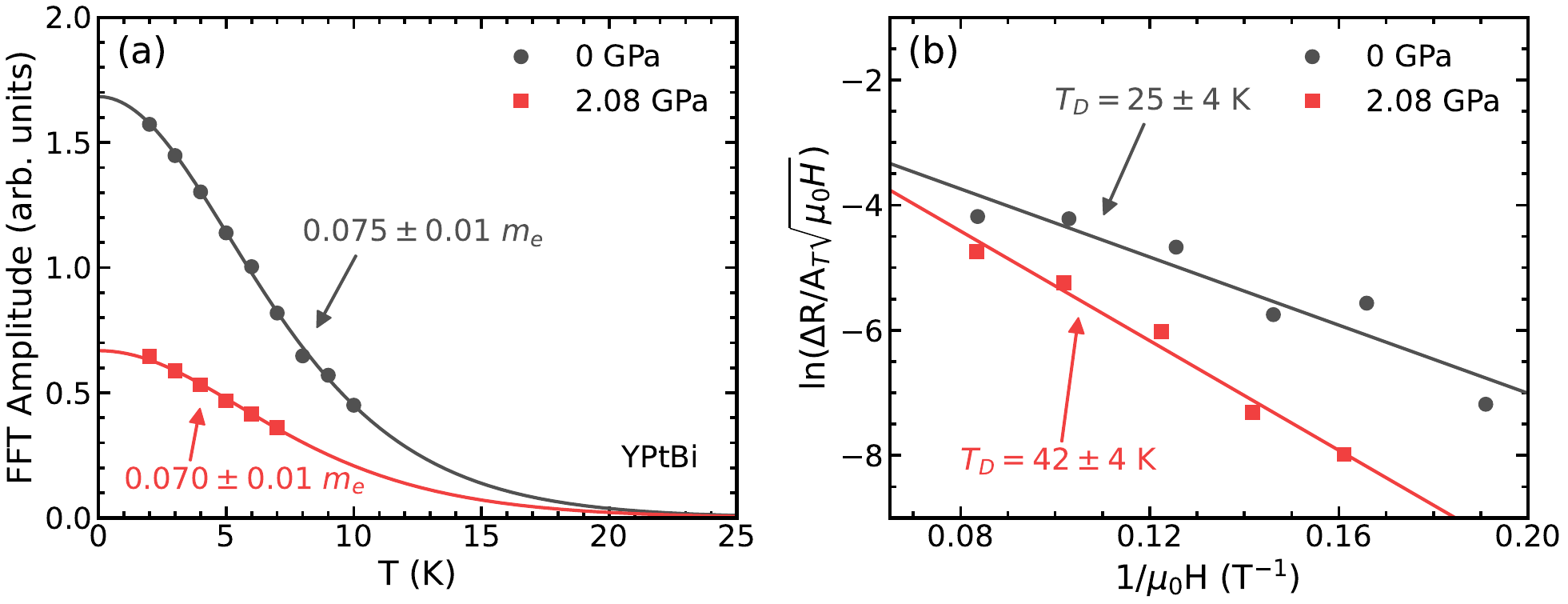}
    \caption{Pressure-dependent effective mass and scattering time in YPtBi. (a) Temperature-dependent QO amplitudes at 0 and 2.08 GPa.
    Symbols represent experimental data while solid lines represent the best fit to the $A_T$ term of the LK formula (Eq. \ref{eq:AT}). (b) Dingle plot of the QOs: $\ln(\Delta R/A_T\sqrt{\mu_0 H})$ as a function of $1/\mu_0H$ at $T = 2$ K. The symbols represent data while the solid lines represent linear fits.}
    \label{fig:QO_LK}
\end{figure*}

Figure \ref{fig:QO_LK}(a) shows temperature-dependent QO amplitudes at 0 and 2.08 GPa as determined via FFT.
The data were then fit to Eq. \ref{eq:AT} to extract $m^*$.
The results of the fits show a very small decrease in $m^*$ with increased pressure, from 0.075 $m_e$ to 0.070 $m_e$ at 0 and 2.08 GPa, respectively.
Fig. \ref{fig:QO_LK}(b) shows a Dingle plot of the data at $T = 2$ K.
By rearranging Eqs. \ref{eq:AT} and \ref{eq:AD}, we find
\begin{equation}
    \ln(A_D) = \ln\bigg(\frac{\Delta R}{A_T \sqrt{\mu_0 H}}\bigg) = -\frac{\alpha T_D}{\mu_0H} + const.
\end{equation}
$T_D$ can then be extracted by fitting the data to a line.
Unlike $m^*$, $T_D$ increases considerably with the application of pressure.
At 0 GPa, $T_D = 25 \pm 4$ K, consistent with \cite{KimYPtBiQuantumOsc}.
This is slightly higher than in other topological systems such as \ce{Bi2Se3} ($T_D=10$ K)\cite{FauqueBi2Se3QO} and \ce{SmB6} ($T_D = 11$-$26$ K) \cite{XiangaSmB6QO}.
At 2.08 GPa, $T_D = 42 \pm 3$ K.
Comparably large values of $T_D$ have been reported in some electron-doped cuprates \cite{HigginsCuprateQO}.

\begin{figure*}[t]
    \centering
    \includegraphics[width=1\linewidth]{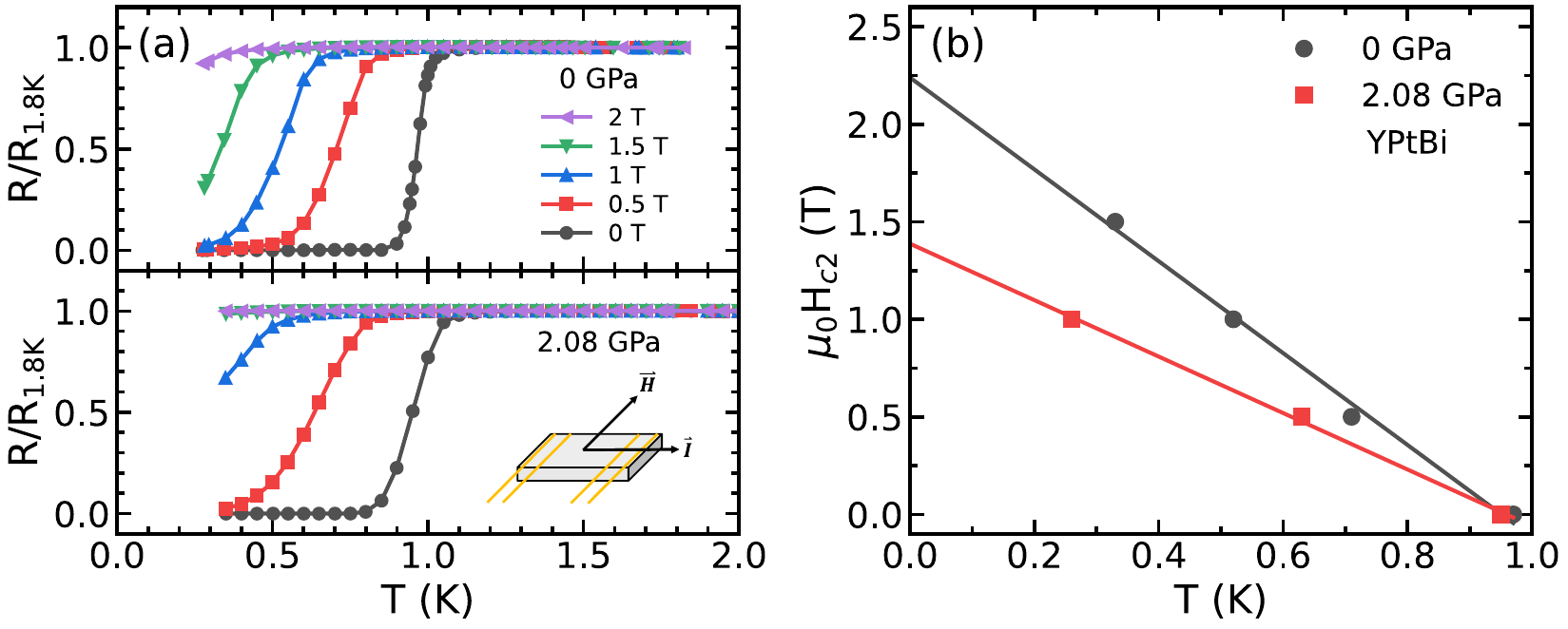}
    \caption{Superconductivity in YPtBi under pressure. (a) The superconducting transition at 0 (top) and 2.08 GPa (bottom). Magnetic field is applied in-plane with and perpendicular to the current. (b) Temperature variation of $H_{c2}$ under pressure. The application of pressure has no appreciable effect on the zero field $T_c$ but broadens the transition and suppresses $H_{c2}.$ }
    \label{fig:Superconductivity}
\end{figure*}

Lastly, we report the effects of pressure on the superconductivity of YPtBi.
Figure \ref{fig:Superconductivity}(a) shows the superconducting transition at 0 and 2.08 GPa, with magnetic field applied in-plane with the current.
We define $T_c$ as the midpoint of the transition, where the resistance equals half its normal state value.
For the partial transition at 2.08 GPa, we extrapolated the transition to half the normal state resistivity. 
In zero field, YPtBi exhibits $T_c \approx 0.95$ K for both pressures but with a broader transition at 2.08 GPa.
At both pressures, we find $H_{c2}$ has a linear temperature dependence within this temperature range (Fig. \ref{fig:Superconductivity}(b)), consistent with earlier data \cite{ButchYPtBi,BayYPtBiPressure}.
Extrapolating to 0 K yields $H_{c2}(0) = 2.24$ T at 0 GPa and $H_{c2}(0) = 1.39$ T at 2.08 GPa, showing a considerable suppression of the upper critical field under pressure.
This can then be used to calculate the superconducting coherence length with $\xi = \sqrt{\frac{\Phi_0}{2\pi H_{c2}}}$, where $\Phi_0$ is the flux quantum.
We find $\xi_{0\ \mathrm{GPa}} = 15.4$ nm and $\xi_{2.08\ \mathrm{GPa}} = 12.1$ nm.

Our observations of the QOs pose an interesting question: why does pressure increase the scattering rate so dramatically but not the frequency or effective mass?
Extrinsic factors such as sample geometry could be responsible; however, the sample was not removed from the pressure cell between subsequent measurements (i.e. it had the same contact geometry, wiring, etc).
Field orientation could also affect the QO amplitude since the amplitude of quantum oscillations in YPtBi shows a strong angle dependence due to the $j$=3/2 quasiparticle character \cite{KimYPtBiQuantumOsc}, but this would require the sample to rotate close to 45$^\circ$, which is unlikely in the pressurized medium of the cell. 
Non-hydrostatic conditions could also lead to increased QO damping due to strain gradients across the sample.
While this is possible, large pressure gradients would likely cause a broadening of the Pb manometer's superconducting transition, which we do not observe.
Furthermore, Daphne oil is known to have good hydrostaticity in this pressure range StaskoPressureMedia, KlotzPressureMedia
Together these indicate the suppression of QO amplitude is an intrinsic effect of pressure.

It is enticing to consider the possibility that pressure is tuning the band inversion in YPtBi.
Band structure calculations by Feng et al. showed that compressing the unit cell weakens the band inversion in the half-Heusler topological semimetals and can even suppress the inversion completely \cite{FengHHBandStruc}.
One thought is that weakened band inversion could lead to an increase in transport scattering as phase space restrictions due to band structure topology lessen.
In principle, this could increase the scattering rate without affecting the size of the Fermi surface or effective mass.
This may also explain the trend towards insulating behavior in the resistivity, which could indicate suppression of the surface conductivity discussed in \cite{KimYPtBiSurface}.
Other possible explanations include structural variations or changes in impurity scattering.
Pressure-induced structural transitions in half-Heusler compounds have been reported previously but at considerably higher pressures \cite{XieCaAuBiPressure}, so this is unlikely in YPtBi.
Likewise, changes in impurity scattering at these low pressures are expected to be minimal since the band structure does not change much.

Lastly we discuss YPtBi superconductivity under pressure.
The insensitivity of the transition to pressure is somewhat surprising.
With the increased scattering rate, one might naively expect a suppression of superconductivity due to stronger pair-breaking effects.
On the other hand, we do observe a nearly 40\% decrease in $H_{c2}$, which is more in line with an increase in orbital pair-breaking.
In either case, this is at odds with the enhancement of superconductivity reported by Bay et al. \cite{BayYPtBiPressure}.
It is worth noting that the decrease in $H_{c2}$ is not explained well by the small changes in effective mass.
We can relate the two via the coherence length with $\xi = \sqrt{\frac{\hbar}{2m^*|\alpha(T)|}}$, where $\alpha(T)$ is one of the Ginzburg-Landau theory    parameters.
Taking the ratio of the coherence lengths, we find that the observed suppression of $H_{c2}$ would require $m^*_{2.08\ \mathrm{GPa}} \approx 0.12\ m_e$, which is considerably larger than the mass obtained from the QOs.
It is also tempting to consider that the suppression of $H_{c2}$ may be related to surface superconductivity, given the in-plane field configuration.
This would imply that the surface-enhanced critical field observed in Ref.~\cite{KimYPtBiSurface} is being suppressed with pressure, also in line with the suppression of the low-temperature saturation in resistivity.
Confirming this would require a more thorough field rotation study.

\section{Conclusion}
To summarize, we performed high-pressure magnetotransport measurements on semi-metallic YPtBi.
Under pressure, we observe a trend towards insulating behavior in the resistivity and no change in the quantum oscillation frequency.
There is also a suppression of the oscillation amplitude that is explained by a large increase in the scattering rate, implying that pressure is tuning the strength of the band inversion.
Lastly, we observe no change in $T_c$, some broadening of the superconducting transition, and a suppression of $H_{c2}$ at 2.08 GPa, in contrast to previous measurements \cite{BayYPtBiPressure}.
Our work situates hydrostatic pressure as a useful tuning parameter to study band structure and superconductivity in this material and likely the other RPtBi and RPdBi \cite{NakajimaRPdBi} half-Heusler compounds.

\section{Acknowledgments}
The authors acknowledge useful discussions with N.P. Butch, K. Rabe, and H. Kim.
Research at the University of Maryland was supported by
the Gordon and Betty Moore Foundation’s EPiQS Initiative Grant No. GBMF9071 (sample preparation),
%the U.S. National Science Foundation Grant No. DMR2303090 (sample preparation), 
%the Air Force Office of Scientific Research Grant No. FA9950-22-1-0023 (sample preparation), 
the Department of Energy, Office of Basic Energy Sciences Award No. DE-SC-0019154 (experimental measurements), 
the NIST Center for Neutron Research, and the Maryland Quantum Materials Center.

\bibliography{YPtBi_Pressure_Paper}

\end{document}